\documentstyle[aps]{revtex}

\begin{document}

\draft

\title{Damping Rate of a Hard Photon in a Relativistic Plasma}
\author{M.H. Thoma}
\address{Institut f\"ur Theoretische Physik, Universit\"at Giessen,\\
35392 Giessen, Germany}
\date{\today}
\maketitle

\begin {abstract}

The damping rate of a hard photon in a hot relativistic QED and QCD plasma
is calculated using the resummation technique by Braaten and Pisarski.

\end{abstract}



\section{Introduction}

Damping rates of hard particles in a relativistic plasma at high temperature
attracted much interest recently
\cite{r1,r2,r3,r4,r5,r6,r7,r8,r9,r10,r11,r12,r13,r14,r15,r16,r17,r18,r19}.
For instance, the damping
rate of a quark in a QCD plasma is related to important quantities of
the quark-gluon plasma, such as mean free paths, thermalization times
\cite{r18,r20}, viscosity \cite{r18,r21}, and stopping power
\cite{r5,r7,r22,r23,r24,r25}. Damping rates
also represent simple tests for methods of the finite temperature field
theory, probing e.g. the infrared behavior \cite{r16} and gauge dependence
\cite{r10,r11,r13,r14,r26,r27,r28} of the methods.

Hard quarks and gluons, i.e. partons with energies $E{\buildrel >\over \sim}
T$, as well as hard electrons and muons in a relativistic  QED or QCD
plasma are damped to lowest order by elastic scattering off the thermal
particles mediated by the exchange of a gauge boson. Owing to the long
range interaction (massless gauge boson) those processes exhibit a
quadratic infrared singularity. The infrared behavior can be improved by
resumming the gauge boson self energy in the "Hard Thermal Loop"
approximation, leading to an effective gauge boson propagator containing
screening effects (Debye screening). This prescription is a special
case of the Braaten-Pisarski resummation technique \cite{r29} which
produces consistent results, i.e. gauge independent results for observables
that are complete to leading order in the coupling constant. The use
of the effective gauge boson propagator, containing the square of the
coupling constant $g^2$ in the denominator, reduces the order of the
damping rates, naively expected to be proportional to $g^4$, to $g^2$,
which was called anomalous (large) damping \cite{r2,r3,r4}. At the same time
the quadratic infrared singularity is replaced by a logarithmic one,
reflecting the absence of static magnetic screening in the effective gauge
boson propagator. (This result led to speculations that these rates
might become finite by higher order effects such as a non-perturbative
magnetic screening mass or self consistent damping \cite{r16} or that these
rates cannot be regarded as direct observables but should be modified by
transport factors \cite{r2,r3,r4,r18}.)

Here we consider the damping of a high energy photon with four momentum
$P=(p_0, {\bf p} )$ and $p=|{\bf p}|\gg T$ in a relativistic plasma of
electrons and positrons or quarks at a temperature much higher than the bare
masses of those particles. In contrast to partons or leptons
the leading order damping mechanism is not caused by elastic scattering since
photons show no self interaction. To lowest order the photons are damped by
Compton scattering and pair creation  processes shown in Fig.1. Neglecting
the fermion masses the corresponding rate suffers from a logarithmic
infrared singularity. However, this divergence is screened by medium
effects again leading to an effective temperature dependent fermion mass
of the order of $gT$ which cuts off the singularity much more effectively
than the bare mass. Since the divergence is only logarithmic utilizing a
bare propagator it disappears completely taking the effective one into
account, analogously to the energy loss of an energetic parton in the
quark-gluon plasma \cite{r7,r22,r24}. Thus the introduction of a magnetic
screening mass, self consistently damped fermion propagators, or transport
factors is unnecessary in the case of photon damping.

Similarly to the energy loss the photon damping rate can be derived
consistently by applying the Braaten-Yuan prescription \cite{r30}. For
this purpose we decompose the rate into a soft contribution, which requires
an effective fermion propagator, and a hard contribution, where bare
propagators are sufficient. The decomposition is accomplished by
introducing a separation scale $\Lambda $ in the momentum of the exchanged
fermion which drops out at the end after adding soft and
hard contributions if $\Lambda $ is restricted by $gT\ll \Lambda \ll T$,
which is possible in the weak coupling limit $g\ll 1$. (In the case of
partons or leptons the hard contribution can be neglected since the
corresponding damping rates are dominated by small momentum transfers.)

The damping rate is defined as the imaginary part of the photon dispersion
relation, $\gamma =-Im \, E(p)$, where the dispersion relation for
a real, i.e. transverse, photon follows from
\begin{equation}
p_0^2-p^2-\Pi _T(p_0, p) = 0,
\label{e1}
\end{equation}
where the transverse part of the photon self energy is given by
\begin{equation}
\Pi _T(p_0,p) = \frac {1}{2}\> \left (\delta _{ij}-\frac {p_ip_j}{p^2}
\right )\> \Pi _{ij} (p_0, p).
\label{e2}
\end{equation}
In the case of no overdamping ($p\ll \gamma $) we get from (\ref{e1})
\begin{equation}
\gamma = -\frac {1}{2p}\> Im\, \Pi_T (p_0=p,p).
\label{e3}
\end{equation}

The damping rate is related to the decay rate $\Gamma _d$ by $\gamma =
\Gamma _d/2$, where $\Gamma _d$ follows from a momentum integration over
matrix elements and distribution functions (see (\ref{e8}) and (\ref{e9}))
\cite{r31}.
We will calculate the soft contribution contribution starting from
(\ref{e3}), whereas the hard contribution starting from the decay rate.

The soft contribution ($k_0$, $k<\Lambda $) follows from the photon self
energy shown in Fig.2, where the blob denotes the effective fermion
propagator containing the one-loop fermion self energy in the Hard
Thermal Loop approximation \cite{r29}. (According to the rules of the
Braaten-Pisarski method we do not need an effective vertex or two effective
fermion propagators at the same time since the external photon is hard.)
Cutting through the blob and a bare propagator we see that
these diagrams correspond to
Compton scattering and pair creation via the exchange of a dressed
(collective) fermion.

First we will consider the damping of a photon in a hot relativistic QED
plasma. The photon self energy shown in Fig.2 is given by
\begin{equation}
\Pi _{\mu \nu} = 2\, i\, e^2\> \int \frac{d^4K}{(2\pi )^4}\> tr\,
[\gamma _\mu\> S(Q)\gamma _\nu \> S^\star (K)],
\label{e4}
\end{equation}
where $S^\star (K)$ indicates the effective fermion propagator, the factor
of 2 comes from adding the both diagrams in Fig.2, and $Q=P-K$.
It is convenient to use
the helicity representation of the fermion propagator given in \cite{r32,r33}.
(The effective electron propagator differs from the quark propagator only
by the effective mass $m_e^2=e^2T^2/8$ compared to $m_q=g^2T^2/6$.)

Evaluating the trace over the gamma matrices we get for the transverse
part of the polarization tensor
\begin{equation}
\Pi _T = 2\, i\, e^2\> \int \frac {d^4K}{(2\pi )^4}\> \left [\frac {1}{D_+(K)}
\left (\frac {1-V}{d_+(Q)}+\frac {1+V}{d_-(Q)}\right ) + \frac  {1}{D_-(K)}
\left (\frac {1+V}{d_+(Q)}+\frac {1-V}{d_-(Q)}\right )\right ],
\label{e5}
\end{equation}
where the functions $D_\pm (K)$ and $d_\pm (Q)$ are given for example
by Kapusta
et al. \cite{r33} and $V=({\bf p}\cdot {\bf k})({\bf p}\cdot {\bf q})/(p^2
qk)$.

Using the imaginary time formalism the integration over $k_0$ in (\ref{e5})
is replaced by a discrete sum, which can be evaluated most easily by
adopting the spectral representation for the fermion propagators
\cite{r32,r33}.
Using furthermore $p\gg T\gg k_0$, $k$ we find
\begin{equation}
Im\, \Pi _T (p_0,p) = -\frac {e^2}{4\pi }\> \int _0^\Lambda dk\> \int _{-k}
^{k} d\omega \> [(k-\omega )\> \beta _+(\omega ,k)+(k+\omega )\> \beta
_-(\omega ,k)],
\label{e6}
\end{equation}
where $\beta _\pm=-Im\, D_\pm^{-1}/\pi$ are the discontinuous parts of the
spectral functions belonging to the effective fermion propagator.
This expression agrees up to some prefactors with the photon production
rate. Thus following Kapusta et al. \cite{r33} we end up with
\begin{equation}
\gamma _{soft} = \frac {\pi }{4}\> \frac {\alpha ^2T^2}{p}\> \ln \frac
{\Lambda ^2}{\pi \alpha T^2}.
\label{e7}
\end{equation}

The hard contribution, for which the momentum of the exchanged fermion is
larger than $\Lambda $, can be calculated most conveniently from the
matrix elements according to Fig.1 of Compton scattering
\begin{eqnarray}
\gamma _{hard}^{comp} = \frac {1}{4p} &\! & \int \frac  {d^3k}{(2\pi )^32k}\>
n_F(k)\> \> \int \frac {d^3p'}{(2\pi )^32p'}\> [1+n_B(p')]\> \int
\frac {d^3k'}{(2\pi )^32k'}\> [1-n_F(k')]\nonumber \\
& \, & (2\pi )^4\> \delta ^4(P+K-P'-K')\> 4\, \langle |{\cal M}|^2\rangle
_{comp}
\label{e8}
\end{eqnarray}
and pair creation
\begin{eqnarray}
\gamma _{hard}^{pair} = \frac {1}{4p} & \! & \int \frac  {d^3k}{(2\pi )^32k}\>
n_B(k)\> \> \int \frac {d^3p'}{(2\pi )^32p'}\> [1-n_F(p')]\> \int
\frac {d^3k'}{(2\pi )^32k'}\> [1-n_F(k')]\nonumber \\
& \, & (2\pi )^4\> \delta ^4(P+K-P'-K')\> 2\, \langle |{\cal M}|^2\rangle
_{pair}.
\label{e9}
\end{eqnarray}
Here momenta with a prime belong to outgoing particles. The factors in
front of the amplitudes, averaged over initial states and
summed over final ones, come from summing over the possible states of the
incoming thermal particle. The matrix elements using Mandelstam variables,
$s=(P+K)^2$, $t=(P-P')^2$, $u=-s-t$, are given by \cite{r34}
\begin{eqnarray}
\langle |{\cal M}|^2\rangle _{comp} & = & -2\, e^4\> \left (\frac {u}{s}
+\frac {s}{u} \right ),\nonumber \\
\langle |{\cal M}|^2\rangle _{pair} & = & 2\, e^4\> \left (\frac {u}{t}
+\frac {t}{u} \right ).
\label{e10}
\end{eqnarray}

In order to evaluate the integrations over the final states in (\ref{e8})
and (\ref{e9}) we assume $p'\gg T$ and $k'\gg T$, which is justified since
$p>p'+k'\gg T$ and the phase space for $p'{\buildrel <\over \sim }T$ or
$k'{\buildrel <\over \sim }T$ is unfavorable \cite{r33}. Thus we may assume
$1+n_B(p')\simeq 1-n_F(p') \simeq 1-n_F(k') \simeq 1$ corresponding to the
Boltzmann approximation for the distribution functions, e.g. $1+n_B(p')
=\exp (p'/T)\, n_B(p')\simeq 1$, also used by Kapusta et al. \cite{r33} in
the case of photon production, where, however, an integration over the initial
states was considered. This approximation simplifies the expressions
(\ref{e8}) and (\ref{e9}) considerably because we can now evaluate the
integrals over $p'$ and $k'$ transforming to the center of mass system
using the Lorentz invariant phase space factor \cite{r34}
\begin{equation}
dQ = (2\pi )^4\> \delta ^4(P+K-P'-K')\> \frac {d^3p'}{(2\pi )^32p'}\>
\frac {d^3k'}{(2\pi )^32k'} =  \frac {dt}{8\pi s}.
\label{e11}
\end{equation}
Let us first consider Compton scattering:
\begin{equation}
\int dQ\> \langle |{\cal M}|^2\rangle _{comp} = \frac {e^4}{4\pi }\>
\left (\ln \frac {s}{\Lambda ^2} + \frac {1}{2}\right ),
\label{e12}
\end{equation}
where $\Lambda ^2\ll s$ cuts off the logarithmic infrared divergence of the
$u$-channel.

Using $s=2pk(1-\hat p\cdot \hat k)$ we obtain after integrating over $d^3k$
\begin{equation}
\gamma _{hard}^{comp} = \frac {\pi }{12}\> \frac {\alpha ^2T^2}{p}\>
\left [\ln \frac {8pT}{\Lambda ^2} + \frac {1}{2} - \gamma + \frac
{\zeta '(2)}{\zeta (2)}\right ],
\label{e13}
\end{equation}
where $\gamma  =0.57722$ is Euler's constant and $\zeta (z)$ is Riemann's
zeta function with $\zeta '(2)/\zeta (2)= -0.56996$ \cite{r7}.

The pair creation contribution gives analogously
\begin{equation}
\gamma _{hard}^{pair} = \frac {\pi }{6}\> \frac {\alpha ^2T^2}{p}\>
\left [\ln \frac {4pT}{\Lambda ^2} - 1 - \gamma + \frac
{\zeta '(2)}{\zeta (2)}\right ].
\label{e14}
\end{equation}
Adding up the (\ref{e13}) and (\ref{e14}) we get the total hard contribution
\begin{equation}
\gamma _{hard} = \frac {\pi }{4}\> \frac {\alpha ^2T^2}{p}\>
\left [\ln \frac {pT}{\Lambda ^2} + \frac {1}{3}\, \ln 128 -
\frac {1}{2} - \gamma + \frac {\zeta '(2)}{\zeta (2)}\right ].
\label{e15}
\end{equation}
Adding up the soft contribution (\ref{e7}) and the hard one (\ref{e15})
the separation scale $\Lambda $ cancels, as expected, yielding the final
result
\begin{equation}
\gamma = \frac {\pi }{4}\> \frac {\alpha ^2T^2}{p}\>
\ln \frac {0.3090\, p}{\alpha T}.
\label{e16}
\end{equation}
Inserting into this expression a typical temperature for a supernova core,
$T=10$ MeV, (neglecting a finite chemical potential) a 100 MeV $\gamma
$-ray has a mean free path $\lambda = \gamma ^{-1}/2$ of 0.40 nm.

In the case of a quark-gluon plasma the photon damping is caused by the
processes of Fig.1, where the outgoing photon for Compton scattering
and one incoming for pair creation has to be replaced by a gluon, leading
to
\begin{equation}
\gamma = \frac {5\pi }{9}\> \frac {\alpha \alpha _sT^2}{p}\>
\ln \frac {0.2317\, p}{\alpha _s T}.
\label{e17}
\end{equation}

The mean free path of a 1 GeV photon in a QGP at $T=200$ MeV, extrapolating
the result (\ref{e17}) obtained in the weak coupling limit $g\ll 1$ to
a realistic $\alpha _s=0.3$, is $\lambda =480$ fm.
This value is much larger than the
dimensions of a fireball in ultrarelativistic heavy ion collisions
confirming that photons may be used as a direct probe of the fireball
\cite{r33,r35}. (Of course, higher order processes such as bremsstrahlung may
contribute to the rate similarly for realistic values of the strong coupling
constant.)

Comparing (\ref{e17}) with the photon production rate $p\, dR/d^3p$
\cite{r33,r36}
we observe that the latter differs only by a factor $4/(2\pi )^3\, \exp
(-p/T)$. The Boltzmann factor reflects the principle of detailed balance
relating the decay rate to the inverse rate $\Gamma _i$ by $\Gamma _d(p)
=\exp (-p/T)\, \Gamma _i(p)$ \cite{r31}. The factor $4/(2\pi )^3$ is just a
matter of definition: $p\, dR/d^3p=2\Gamma _i/(2\pi )^3$ and $\gamma =
\Gamma _d/2$. This also shows that the hard part of the photon production
rate \cite{r33,r37} can be calculated much easier by considering the decay
instead of the production process.

\begin{figure}
\caption{Lowest order Feynman diagrams for photon damping: (a) Compton
scattering, (b) pair creation.}
\end{figure}

\begin{figure}
\caption{Photon self energy diagrams containing an effective fermion
propagator denoted by a blob.}
\end{figure}

\end{document}